\documentclass[letter]{aa}

\usepackage{txfonts}
\usepackage{graphicx}

\begin{document}

   \title{Evidence of the inhomogeneity of the stellar population in the differentially
reddened globular cluster NGC 3201.
   \thanks{Based on observations with the 1.3 m Warsaw telescope at Las Campanas
Observatory}}

      \author{V. Kravtsov\inst{1,2}
          \and
              G. Alca\'ino\inst{3}
          \and
              G. Marconi\inst{4}
           \and
              F. Alvarado\inst{3}
              }

\offprints{V. Kravtsov}

   \institute{Instituto de Astronom\'ia, Universidad Cat\'olica del Norte,
              Avenida Angamos 0610, Antofagasta, Chile\\
              \email{vkravtsov@ucn.cl}
            \and
              Sternberg Astronomical Institute, University Avenue 13,
              119899 Moscow, Russia\\
            \and
              Isaac Newton Institute of Chile, Ministerio de Educaci\'on de Chile,
              Casilla 8-9, Correo 9, Santiago, Chile\\
              \email{inewton@terra.cl, falvarad@eso.org}
            \and
              ESO - European Southern Observatory, Alonso de Cordova 3107, Vitacura,
              Santiago, Chile\\
              \email{gmarconi@eso.org}
             }

   \date{Received xxxxx / Accepted xxxxx}

   \abstract
{} {We report on evidence of the inhomogeneity (multiplicity) of the
stellar population in the Galactic globular cluster (GC) NGC 3201,
which is irregularly reddened across its face.} {We carried out a
more detailed and careful analysis of our recently published new
multi-color photometry in a wide field of the cluster with
particular emphasis on the $U$ band.} {Using the photometric data
corrected for differential reddening, we found for the first time
two key signs of the inhomogeneity in the cluster's stellar
population and of its radial variation in the GC. These are (1) an
obvious trend in the color-position diagram, based on the $(U-B)$
color-index, of red giant branch (RGB) stars, which shows that the
farther from the cluster's center, the bluer on average the $(U-B)$
color of the stars is; and (2) the dependence of the radial
distribution of sub-giant branch (SGB) stars in the cluster on their
$U$ magnitude, where brighter stars are less centrally concentrated
than their fainter counterparts at a confidence level varying
between 99.2\% and 99.9\% depending on the color-index used to
select the stars. The same effects were recently found by us in the
GC NGC 1261. However, contrary to NGC 1261, we are not able to
unambiguously suggest which of the sub-populations of SGB/RGB stars
can be the progenitor of blue and red horizontal branch stars of the
cluster. Apart from M4, NGC 3201 is another GC very probably with an
inhomogeneous stellar population, which has essentially lower mass
than the most massive Galactic GCs where multiple stellar
populations were unambiguously detected for the first time.} {}


\maketitle

\section{Introduction}
\label{introduc}

The southern Galactic globular cluster (GC) NGC 3201, known not only
by its peculiar kinematic characteristics but also by irregular
differential reddening across its face, was the subject of our
recent study (Kravtsov et al. \cite{kravtsovetal09}) based on a new
multi-color photometry in a 14$\arcmin$x14$\arcmin$ field of the GC.
In that study, where we primarily dealt with some aspects of the
properties and characteristics of the cluster stellar population, we
also allowed a possible spread in the population, but did not
examine it. In a later more detailed analysis of the same data, we
were able to find not only apparent manifestations, but also
stronger and more objective evidence of the inhomogeneity in the
stellar population and of its radial variation in the cluster. The
present letter reports on these findings in detail. The obtained
results contribute more to our past (Alca\'ino et al.
\cite{alcainoetal99}) and recent (Kravtsov et al.
\cite{kravtsovetal10}) studies of the inhomogeneity (multiplicity)
of the stellar populations in the populous Large Magellanic Cloud
(LMC) cluster NGC 1978 and Galactic GC NGC 1261, respectively and to
the rapidly growing body of photometric and spectroscopic evidence
about multiple stellar populations in both Magellanic Clouds star
clusters (e.g., Mackey et al. \cite{mackeyetal08}; Milone et al.
\cite{milonetal09a}; and references therein) and Galactic GCs (some
relevant publications are referred to elsewhere in the paper).

\section{The used photometric data and the corrections applied to them}
\label{photdat}

In the present study, we are relying on our recent multi-color
photometry in $UBVI$ of more than 12\,000 stars in a
14$\arcmin$x14$\arcmin$ cluster field, reaching below the turnoff
point in all passbands. For a description of the photometric data
used here (and publicly available in electronic form) and of the
typical r.m.s errors in each passband, see Kravtsov et al.
(\cite{kravtsovetal09}). In the same paper, we also described in
detail our study of the variation of reddening in the observed
cluster field and listed the estimated corrections corresponding to
areas with conditionally defined grades of reddening. These
corrections were estimated separately for the $U$ and $V$
magnitudes, as well as for the $(B-V)$, $(V-I)$, and $(B-I)$ colors,
and typical uncertainties were quoted. As for the corrections for
the $(U-B)$, they were supplementary calculated in the present study
from those originally obtained for the $(B-V)$ color by applying the
following relation between the color-excesses in both colors:
$E_{U-B}$ = 0.72$E_{B-V}$. In the analysis presented below, we used
photometric data and the cluster's CMDs corrected for differential
reddening.

In addition to its variable reddening, the cluster field is also
populated by the large number of field stars, which are fairly
numerous in the region of the lower red giant branch (RGB) and
sub-giant branch (SGB). A decontamination procedure was applied to
the photometry used. For more detail, interested readers are
referred to our original paper on photometry of NGC 3201.

\section{The sub-giant branch in the $U$ magnitude}
\label{sgb}

\begin{figure}
   \centering
   \includegraphics[angle=-90,width=8.8cm]{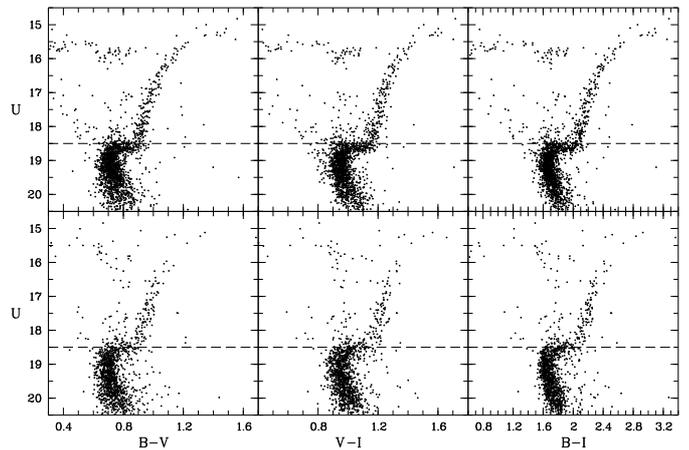}
         \caption{The $U$-based CMDs with different color-indexes, corrected
for differential reddening and (partially) rid of field stars,
demonstrating the systematic variation of the position of the SGB in
the $U$ with radial distance from the center of NGC 3201. Upper and
lower panels show the CMDs of inner ($1\farcm35 < R < 2\farcm70$)
and outer ($R > 3\farcm40$) regions of the cluster, respectively.
The dashed horizontal line is given as a reference to facilitating a
comparison of the SGB in the CMDs.}
         \label{cmd}
   \end{figure}

In two rows of panels of Fig.~\ref{cmd}, we demonstrate the apparent
radial variation of the $U$-level of the SGB of NGC 3201 in the
$U$-based CMDs with different color-indexes. The CMDs are corrected
for differential reddening and decontaminated of the majority of
field stars in the region of the lower RGB and of the SGB. The CMDs
shown in the upper and lower panels correspond to the inner
($1\farcm35 < R < 2\farcm70$) and the outer ($R > 3\farcm4$) regions
of the cluster, respectively. One can see that at least the
$U$-level and perhaps the slope of the SGB vary with the distance
from the center of NGC 3201, the SGB being systematically brighter
(and perhaps steeper) in the outer part of the cluster The dashed
horizontal line is drawn at the same $U$ magnitude in both rows of
panels as a reference line to facilitate a comparison of the SGB
level in the CMDs of the two parts in the cluster.

The qualitatively demonstrated radial variations of the distribution
of stars on the SGB in the $U$-based CMDs are supported by more
objective indicators based on quantitative estimates. They were
obtained in the following way. We first isolated a sample of stars
most probably belonging to the SGB. The $U$-based cluster CMD with
the $(B-I)$ color-index was initially used to this goal. This
color-index provides the largest separation between the turnoff (TO)
point and the lower RGB, as compared with other available
color-indices corrected for differential reddening. The central part
of the cluster was not excluded from the analysis, thanks to only a
weak crowding effect in it. Given a slight apparent slope of the SGB
in the $U$-$(B-I)$ CMD, it was taken into account in isolating SGB
stars. They were selected in the color range $\Delta (B-I) =$ 0.25
($1.75 < B-I < 2.00$) and the magnitude range $\Delta U =$ 0.4 mag
between the accepted upper and lower borders of the branch, defined
by two envelope lines, $U$ = -0.194$(B-I)$ + 18.75 and  $U$ =
-0.194$(B-I)$ + 19.15. The total number of the selected stars is
415. We then divided the obtained sample of the SGB stars by three
sub-samples: (1) a sub-sample of 115 brightest SGB stars in the
magnitude range $\Delta U =$ 0.15 mag; (2) a sub-sample of 178
faintest SGB stars also confined in the magnitude range $\Delta U =$
0.15 mag; (3) a sub-sample of 122 SGB stars falling in the magnitude
range $\Delta U =$ 0.10 mag in between the two extreme ranges. Then
we also similarly selected SGB stars by relying independently on
each of the two additional color-indices, $(B-V)$ and $(V-I)$, in
their ranges $\Delta (B-V) =$ 0.12 ($0.75 < B-V < 0.87$) and $\Delta
(V-I) =$ 0.13 ($1.00 < V-I < 1.13$), respectively. A total magnitude
range of the SGB, $\Delta U =$ 0.4 mag and a division of it into
three parts yields the same as in the $U$-$(B-I)$ CMD, with the same
$U$ magnitudes defining the boundaries of these parts at the
extremes of each color range as was accepted for the SGB in the
three color-indices. Note that although the color-indices under
consideration are not independent (i.e., each of them is a
combination of two other color-indices), the corrections for
differential reddening were originally made independently in each of
them. The number of stars falling in the fainter, intermediate, and
brighter boxes are 188, 130, and 117 respectively in the $U$-$(B-V)$
CMD. In turn, the boxes of the same sequence of the brightness
levels in the $U$-$(V-I)$ CMD contain 165, 125, and 114 stars,
respectively. For comparison purposes, we also isolated SGB stars in
the $V$ based CMD, namely $V$-$(V-I)$. The SGB is apparently
slightly narrower in the $V$ magnitude, i.e. $\Delta V =$ 0.35 mag
as compared to $\Delta U =$ 0.4 mag. The color range is the same as
in the $U$-$(V-I)$ CMD. The accepted upper and lower borders of the
SGB in the $V$-$(V-I)$ were defined by two envelope lines, $V$ =
-1.923$(V-I)$ + 19.87 and $V$ = -1.923$(V-I)$ + 19.52. Taken the
narrower SGB in the $V$ magnitude into account, we selected
sub-samples of the brightest and faintest SGB stars in slightly
reduced magnitude ranges than in the $U$ magnitude, namely $\Delta V
=$ 0.13 mag (98 stars) and $\Delta V =$ 0.12 mag (136 stars),
respectively. The magnitude range of the box containing the
sub-sample of SGB stars of intermediate brightness (131 items) was
kept unchanged.

\begin{figure*}
\centerline{
\includegraphics[clip=,angle=-90,width=4.5 cm]{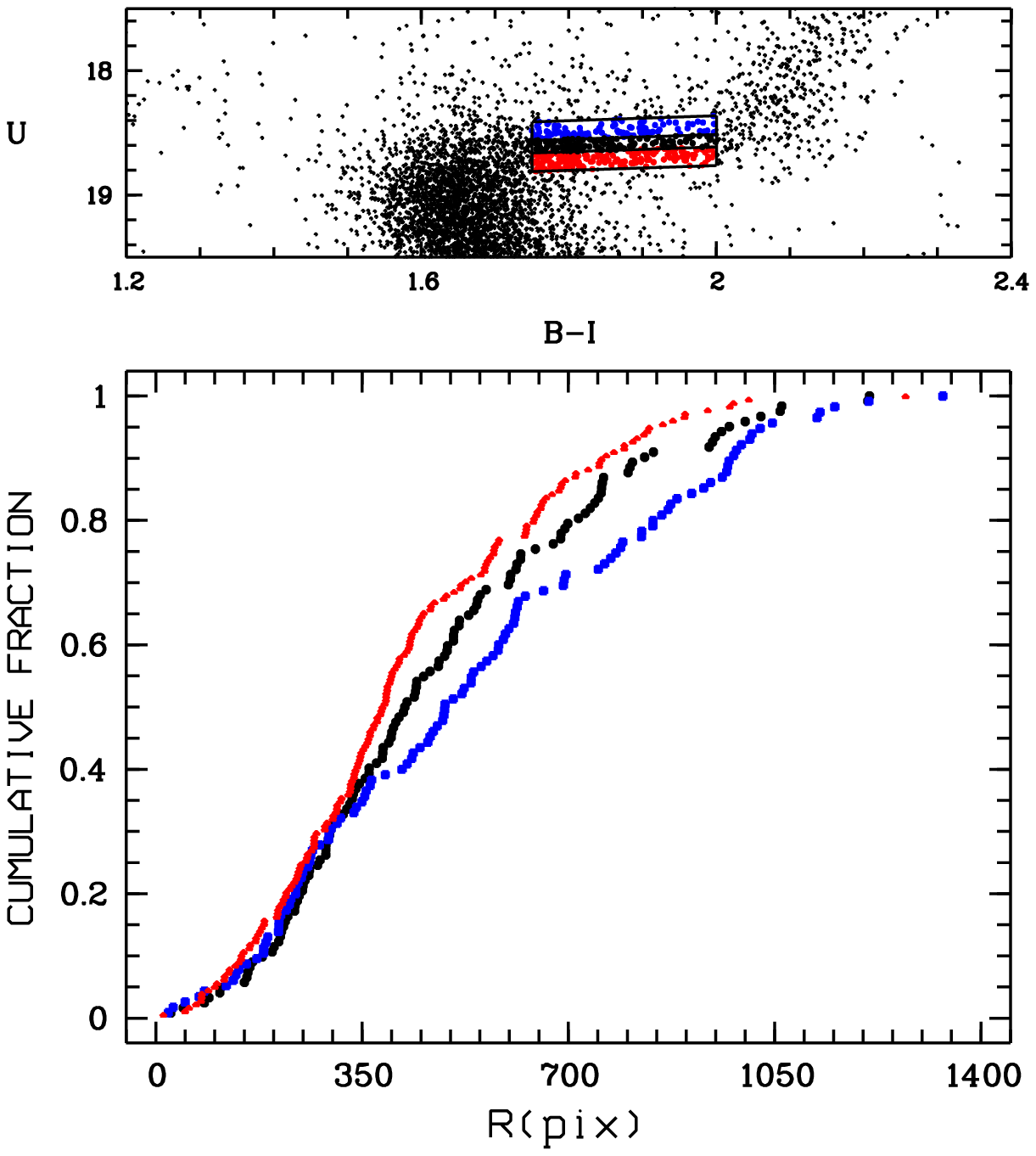}
\includegraphics[clip=,angle=-90,width=4.5 cm]{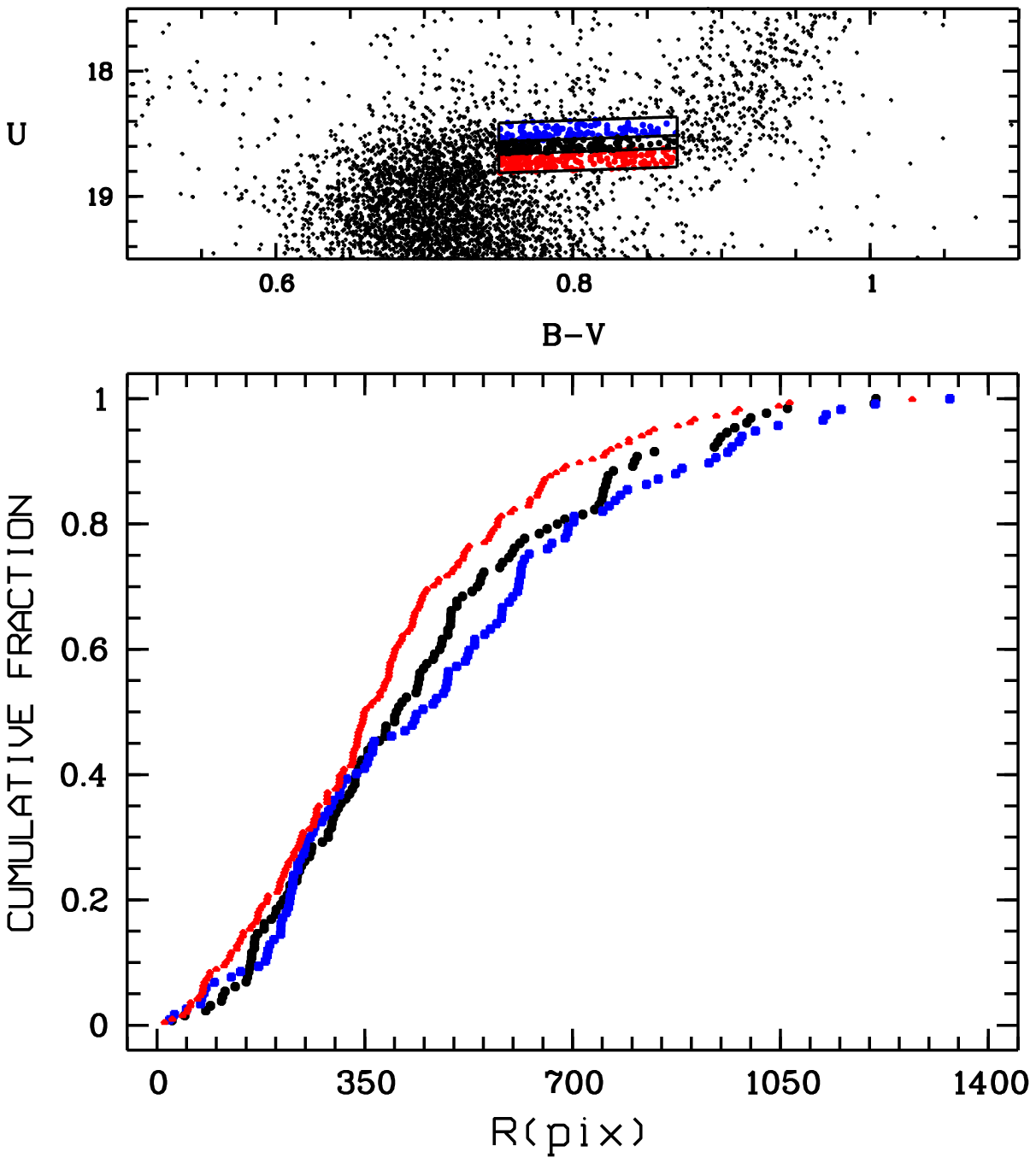}
\includegraphics[clip=,angle=-90,width=4.5 cm]{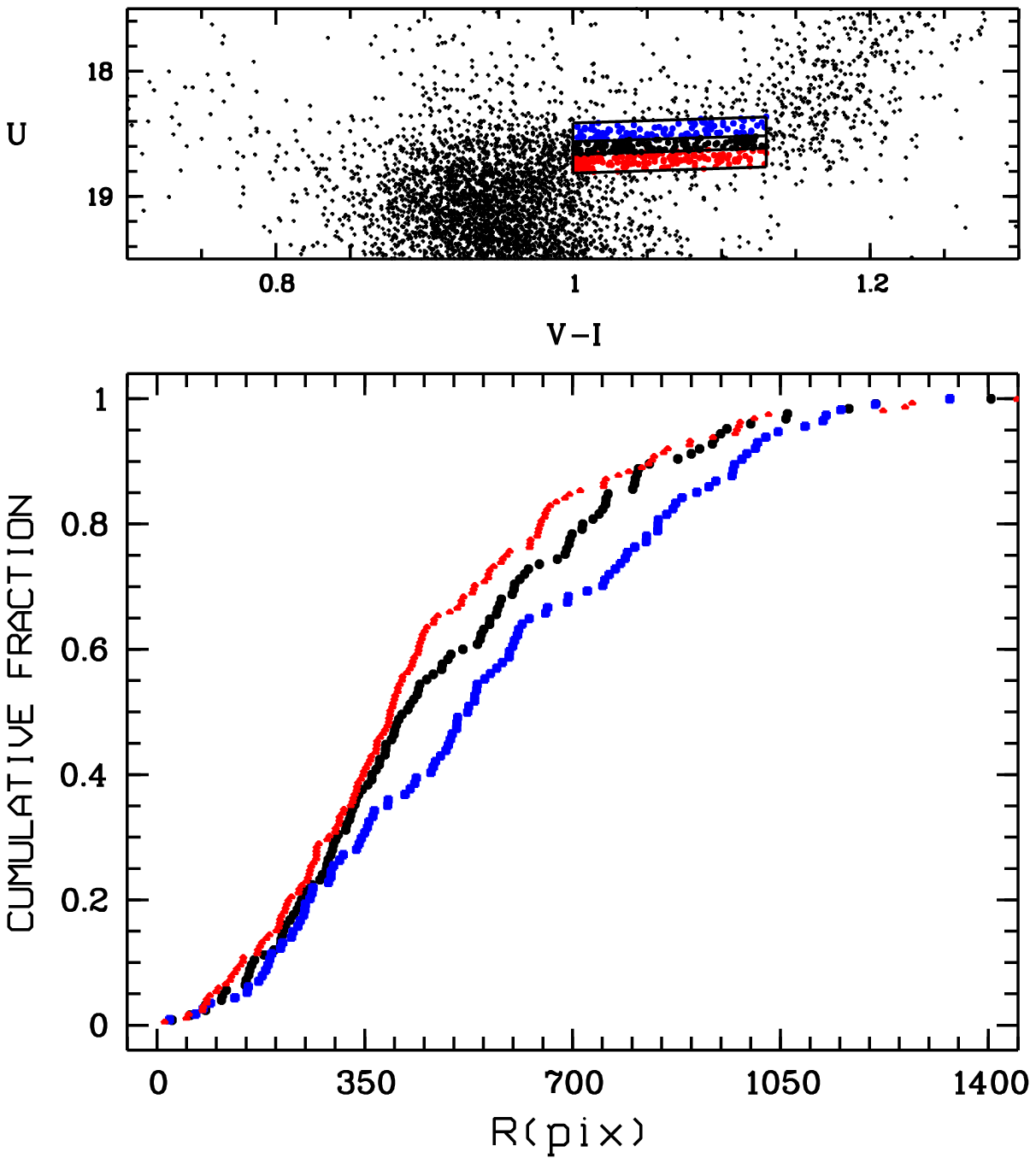}
\includegraphics[clip=,angle=-90,width=4.5 cm]{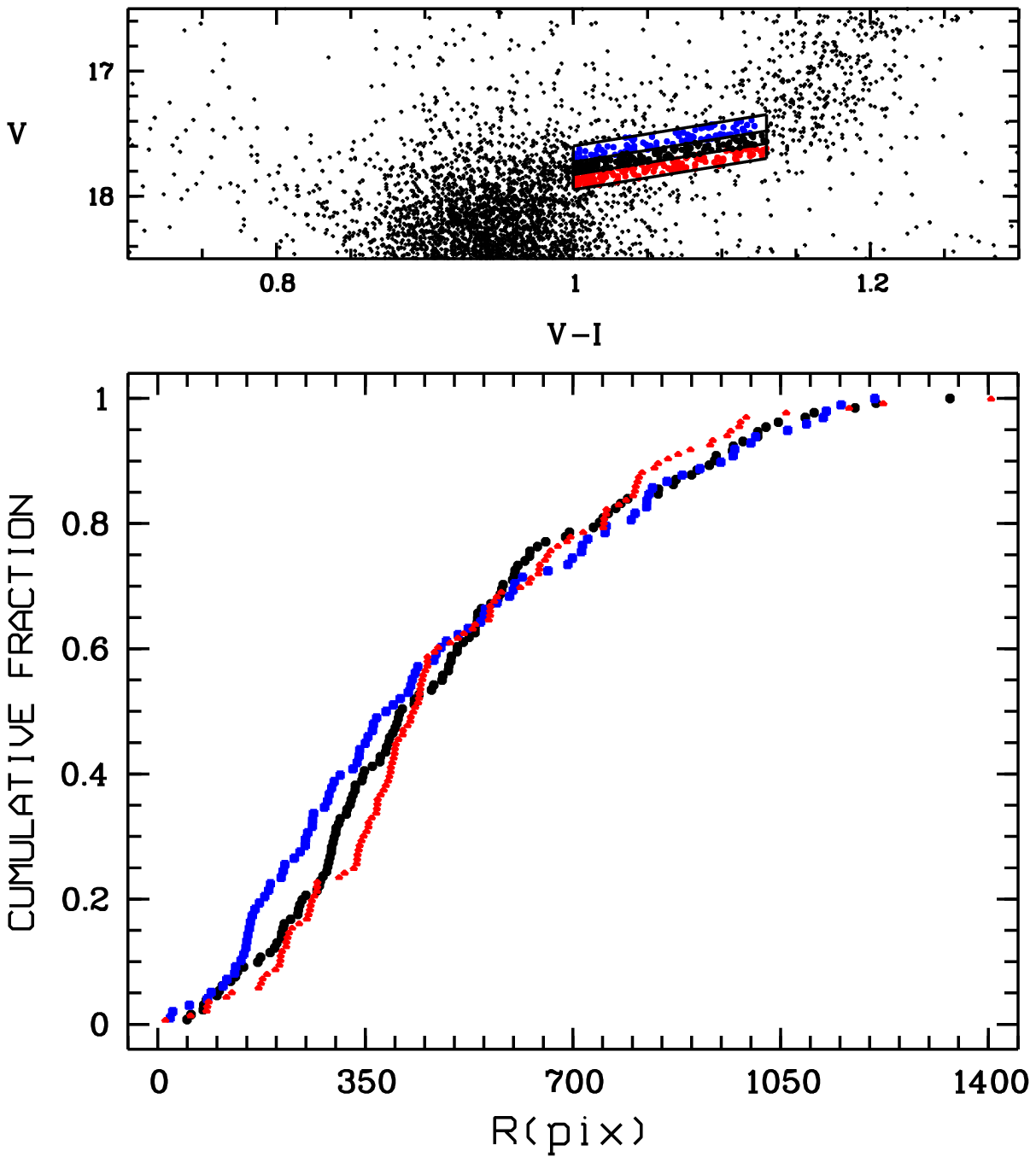}
} \caption{Upper panels show three sub-samples of the SGB stars
isolated in three magnitude ranges: (1) in the $U$-based CMDs of NGC
3201 with different color-indices, and additionally (2) in the
$V$-$(V-I)$ CMD, see text for details. All the CMDs are corrected
for differential reddening. Lower panels: a comparison of the
cumulative radial distributions of the three sub-samples of SGB
stars; red filled triangles, black filled circles, and blue filled
squares denote the sub-samples of the SGB stars with progressively
increasing brightness in the given passbands. In all the panels, the
three sub-samples are shown with symbols of the same color.}
         \label{sgbvar}
   \end{figure*}

The selection boxes are drawn by the solid lines in the CMDs in the
upper panel of Fig.~\ref{sgbvar}, where the corresponding selected
stars of the three sub-samples with progressively increasing
brightness are denoted by red, black, and blue filled circles. The
lower panels of Fig.~\ref{sgbvar} show the cumulative radial
distributions of these sub-samples of SGB stars. As for the stars
isolated in the three $U$ based CMDs, one can see that the faintest
sub-giants denoted by (red) filled triangles are more centrally
concentrated than their counterparts of intermediate brightness in
the $U$ (black filled circles), and obviously yet more centrally
concentrated than the brightest SGB stars (blue filled squares),
irrespective of the color-index used for the selection of the SGB
stars. This apparent difference between the distributions is
supported by a quantitative estimate based on a Kolmogorov-Smirnov
test: the difference between the radial distributions of the
brightest and faintest SGB stars is statistically significant at
99.8\%, 99.2\%, and 99.9\% confidence levels for the selection made
in the $U$-$(B-I)$, $U$-$(B-V)$, and $U$-$(V-I)$ CMDs, respectively.
However, the radial distributions of the sub-samples of SGB stars
with different brightness in the $V$ magnitude are apparently in
contrast to those differing by the $U$ magnitude. In particular, the
brightest SGB stars in the $V$ passband are even somewhat more
centrally concentrated in the central part of NGC 3201 than their
faintest counterparts. Moreover, the radial distributions of the two
sub-samples of SGB stars are different at 92.1\% confidence level.
Such a difference is, strictly speaking, statistically
insignificant, because the estimated confidence level is less than
95.0\%.

The above result may imply the following. Even if the dependence of
the radial distribution of SGB stars on their brightness in the $V$
magnitude (at a given color) really exists and has the same meaning
as the dependence in the $U$ magnitude, it is presumably less
obvious, so that the errors of the applied corrections for
differential reddening combined with the errors of the photometry
itself are able to reduce or even distort the dependence.

The high level of statistical significance of the dependence of the
SGB stars' radial distribution in NGC 3201 on their $U$ magnitude is
strong evidence of the inhomogeneity ("multiplicity") of the cluster
stellar population. This high level was also found by us in another
Galactic GC, NGC 1261. Here we note again that the inhomogeneity of
the population of SGB stars resulting from the revealed dependence
in both GCs resembles a combination of two effects: (1) the split of
the SGB into two components revealed by Milone et al.
(\cite{milonetal08}) in the GC NGC 1851 and (2) the different radial
distribution of stars belonging to the brighter and fainter
components of the SGB, found by Zoccali et al. \cite{zoccalietal09})
in the same GC, i.e. the brighter SGB stars were found to be less
centrally concentrated and to extend to much larger radial distances
in NGC 1851. Note however that Milone et al. (\cite{milonetal09b})
do not support this finding.

As for the revealed differences regarding SGB stars in NGC 3201, one
cannot draw any definite conclusion about whether or not they are of
a discrete or a continuous character.

\section{The red giant branch and its color-position diagram}
\label{rgb}

\begin{figure}
  \centering
 \includegraphics[angle=-90,width=5.5cm]{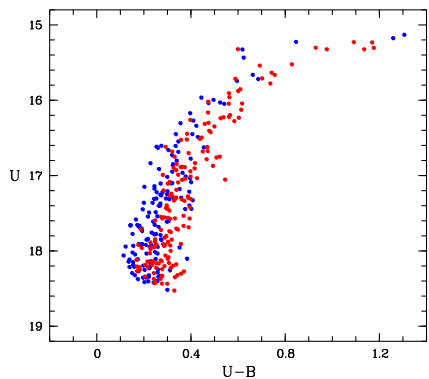}
   \caption{Comparison of the location in the $U$-$(U-B)$ CMD
of RGB stars from the inner (red dots, $1\farcm35 < R < 2\farcm70$)
and outer (blue dots, $3\farcm4 < R < 6\farcm3$) regions of NGC
3201.}
         \label{rgbposit}
   \end{figure}

The obtained evidence of the probable inhomogeneity of SGB stars in
NGC 3201 is reinforced by the inhomogeneity of RGB stars and its
dependence on the radial distance in the cluster. For our analysis
of the RGB of NGC 3201, we used a ready-made sample of the most
probable RGB stars. Their selection was done previously and is
described in detail in Kravtsov et al. \cite{kravtsovetal09}. Based
on this sample we found as in the RGB of NGC 1261 a systematically
different location in the $U$-$(U-B)$ diagram of RGB stars situated
at different radial distances from the cluster center, where the
portion of stars bluer in the $(U-B)$ systematically increased
towards the cluster outskirts. In Fig.~\ref{rgbposit} we compare in
the $U$-$(U-B)$ diagram the position of two sub-samples of RGB stars
belonging to two areas at different mean radial distances from the
cluster center, namely in the radial ranges $1\farcm35 < R <
2\farcm70$ and $3\farcm4 < R < 6\farcm3$. It is evident that RGB
stars in the outer region (blue dots) are systematically bluer than
in the inner one.

\begin{figure}
  \centering
 \includegraphics[angle=-90,width=6.7cm]{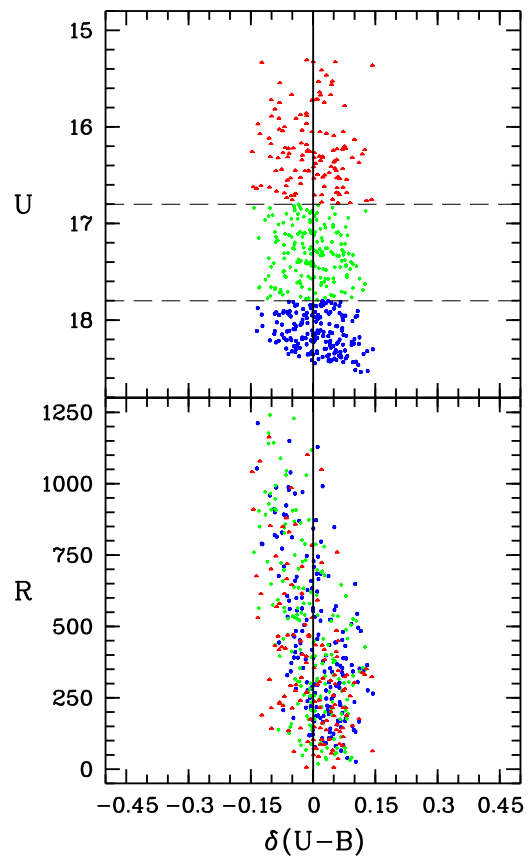}
   \caption{The trend of the $(U-B)$ color of RGB stars with their radial
distance from the center of NGC 3201. Upper panel: the linearized
RGB for the color $(U-B)$ corrected for differential reddening. A
dozen of the brightest stars of the RGB ($U < 15.3$) were rejected.
For demonstration purposes, the RGB was arbitrarily divided into
three magnitude intervals of comparable samples of stars, marked by
dashed lines, where stars are denoted by different symbols. The
lower panel shows the color-position diagram of the same RGB stars.}
         \label{cpd}
   \end{figure}

To secure more convincing evidence and more detailed behavior of the
suggested dependence of the $(U-B)$ color of RGB stars on their
radial distance, R, from the center of NGC 3201, we obtained the
so-called color-position diagram (CPD) of the RGB. Note that this
diagram was used for the first time in our study of the stellar
population in the LMC populous star cluster NGC 1978 (Alca\'ino et
al. \cite{alcainoetal99}). First of all, we linearized the RGB in
the $U$-$(U-B)$ plane by relying on the ready-made sample of the
most probable RGB stars. Since the brightest part of the RGB, formed
by a dozen of stars with $U < 15.3$, is quasi-horizontal in the
$U$-based CMDs, we rejected these stars. We fitted the mean locus of
the RGB with a polynomial and subtracted for each star the color of
the mean locus at its luminosity level from the star's color-index.
We left nearly all stars of the initial sample. Only a handful of
stars with deviations, $\delta(U-B)$, from the mean locus exceeding
$\pm$ 0.15 mag in the total range of the RGB in the $U$ magnitude
were rejected. Note that the number of rejected stars is so small
that this is rather a formal procedure with negligible impact. The
linearized RGB is shown in the $U$-$\delta(U-B)$ plane in the upper
panel of Fig.~\ref{cpd}. With the purpose explained below, the
linearized RGB was arbitrarily divided by three magnitude ranges
(marked by the dashed lines) with comparable samples of stars,
namely: 189 faint stars with $U > 17.8$; 185 stars with intermediate
brightness, $16.8 < U < 17.8$; 113 brighter stars with $U < 16.8$.
The final step is plotting the dependence between the deviations
$\delta(U-B)$ of the RGB stars and their radial distances from the
center of NGC 3201.

The obtained CPD shown in the lower panel of Fig.~\ref{cpd} is not
only another representation of the demonstrated photometric
inhomogeneity of RGB stars in NGC 3201, but also a dependence
containing additional information. It is now evident beyond doubt
that the $(U-B)$ color does get bluer, i.e. the $\delta(U-B)$
becomes more negative with increasing radial distance from the
cluster center. It is also obvious that all the three sub-samples of
RGB stars follow the same trend. Note yet that the CPD assumes the
trend apparently reveals itself beginning at $R \approx 350 - 400$
pixels ($2\farcm35 - 2\farcm70$) and there is no obvious trend
within this radial distance, where the bulk of stars have positive
deviations $\delta(U-B)$. Interestingly enough, this radius is
virtually equal to the cluster half-mass radius, $R_h = 2\farcm68$
(Harris \cite{harris96}). In this connection it should be also noted
that the difference in radial distribution among SGB stars arises
approximately at the same radius.

Concerning the physical reasons that can probably be responsible for
the revealed inhomogeneity, we refer to recently obtained results by
Marino et al. (\cite{marinoetal08}) on stellar population in GC M4.
Based on spectroscopy of a sample of 105 stars in this GC they found
a dichotomy in Na abundance and argued that it must be associated
with a CN bimodality. From photometry of the same stars Marino et
al. (\cite{marinoetal08}) showed that the CN-weak red giants with a
lower content of Na are on average systematically bluer, by $\Delta
(U-B) = 0.17$ in the $U$-$(U-B)$ CMD, than their CN-strong
counterparts with a higher content of Na. This photometric effect is
very similar to that revealed by us in NGC 3201 and in NGC 1261.
From the CPD we evaluate the mean separation in the $(U-B)$ color
between "blue" and "red" RGB stars, assuming this bi-modality in
their distribution in the $(U-B)$ color. It is apparently around
$\Delta (U-B) \sim 0.12$, which is many times larger than the
possible error in the $(U-B)$ color (a few hundredth of magnitudes
in the worse case) caused by the uncertainty in local fluctuations
of reddening in the cluster field under consideration. It agrees
well with our estimate made for NGC 1261 and is on the same order of
magnitude as the above-mentioned difference between the "redder" and
"bluer" giants in M4. A tentative first explanation is to accept the
same reason behind the segregation of RGB stars in the color (U-B)
in NGC 3201 as in M4.

In NGC 3201, other than in NGC 1261, we are not able to suggest an
unambiguous association between the discussed sub-populations of
SGB/RGB stars and those belonging to the blue and red horizontal
branch (BHB, RHB).

\section{Conclusions}

Based on a more detailed analysis of a new multi-color photometry in
an extended field of the differentially reddened GC NGC 3201, we
found the following signs of the inhomogeneity (multiplicity) of the
cluster's stellar population. First, there is an obvious dependence
of the radial distribution of SGB stars in the cluster on their $U$
magnitude: brighter stars are less centrally concentrated than their
fainter counterparts at a confidence level fluctuating above 99.2\%
in relation to the color-index of CMD relied on to isolate SGB
stars. Second, RGB stars exhibit a systematically different location
in both the $U$-$(U-B)$ CMD and the $R$-$\delta(U-B)$ color-position
diagram at different radial distances from the cluster center: the
proportion of stars bluer in the $(U-B)$ increases towards the
cluster outskirts. We note (1) the same kind of photometric
inhomogeneity of RGB and SGB stars in NGC 3201 and in another
Galactic GC, NGC 1261 (Kravtsov et al. \cite{kravtsovetal10}), and
also (2) a very similar radial trend in both GCs.

Finally, it is worth mentioning that NGC 3201 after M4, is the
second non-massive Galactic GC that very probably has an
inhomogeneous stellar population.

\begin{acknowledgements}

We thank the anonymous referee for useful comments that improved the
manuscript.

\end{acknowledgements}

\end{document}